\documentclass[aps,prl,reprint,superscriptaddress]{revtex4-1}
\usepackage{amsmath,amssymb} 
\usepackage{makeidx}
\usepackage{amsfonts}
\usepackage[colorlinks,hyperindex]{hyperref}
\hypersetup
{
	colorlinks,%
	citecolor=black,%
	linkcolor=black,%
	urlcolor=black,%
}
\usepackage[normalem]{ulem}
\newcommand\numberthis{\addtocounter{equation}{1}\tag{\theequation}}
\usepackage[dvips]{graphicx}

\newcommand{\comment}[1]{} 

\newcommand{\added}[1]{{#1}}
\newcommand{\del}[1]{}%{{\color{red}{\sout{#1}}}}

\newcommand{\diag}{\operatorname{diag}}

\newcommand{\D}{\textrm{d}}
\makeindex

\begin{document}

\title{Turbulent mixing: matching real flows to Kraichnan flows}

\author{Siim Ainsaar}
\affiliation{Institute of Cybernetics, Tallinn University of Technology}	
\affiliation{Institute of Physics, University of Tartu}
\author{Mihkel Kree}
\affiliation{Institute of Cybernetics, Tallinn University of Technology}	

\author{Jaan Kalda}
\email{kalda@ioc.ee}
\affiliation{Institute of Cybernetics, Tallinn University of Technology}

\date{\today}

\begin{abstract}
Majority of theoretical results regarding turbulent mixing are based on the 
model of ideal flows with zero correlation time. We
discuss the reasons why such results may fail for real flows and develop a scheme which 
makes it possible to match real flows to ideal flows. 
In particular we introduce the concept of mixing dimension of flows which can take fractional values.
For real incompressible flows, the mixing dimension exceeds the topological dimension; this 
leads to a local inhomogeneity of mixing  --- a phenomenon which is not observed for ideal flows and 
has profound implications, for instance impacting the rate of bimolecular reactions in turbulent flows.
\added{Finally, we build a model of compressible flows which
reproduces the anomalous Lyapunov exponent values observed 
for time-correlated flows by Boffetta et al (2004), and provide a qualitative explanation of this phenomenon.}
\del{In the case of compressible flows, we make additional model assumptions which allow us to explain the 
anomalous Lyapunov exponent values observed for time-correlated flows by Boffetta et al (2004).}
\end{abstract}

\maketitle

Turbulent mixing affects us in many ways, for instance via weather and Solar storms. 
Here we assume that the fields which are being mixed are passive, i.e.\
do not affect the statistical properties of the underlying velocity field.  
Examples of fairly passive fields include
fluid temperature, concentration of  pollutants, nutrients, etc., and
weak magnetic fields at a linear stage of magnetic dynamos;
more examples can be found in reviews 
\cite{Falk,Sreenivasan,Warhaft,Dimotakis,Shraiman,Toschi,Grabowski,Brandenburg}. 

\del{Dominating }
\del{Majority of the }
Analytical studies of turbulent mixing
have been \added{mostly} based on the Kraichnan's model \cite{Kraich}, which assumes 
the velocity field to be $\delta$-correlated in time (with vanishing correlation time). 
\del{The success of this approach is explained by the fact that the }
\added{This approach has been successful because}
properties of turbulently mixed fields depend almost exclusively on the 
stretching statistics of the material elements (point pair separations, 
lines, surface areas, volumes) as a function of time; hence, 
as long as Kraichnan flows match the stretching statistics of real flows, 
the predictions based on \added{them will}
\del{the Kraichnan's model were thought to  }
be \del{fairly} accurate.
\del{While it has been acknowledged that Kraichnan flows do not match the exponential tails of the material elements' stretching statistics (which affects the structure function scaling) \cite{Shraiman}, 
simulations show that the mismatch can be more severe than that: }
\added{Therefore, the finding that} for compressible flows, e.g.\ at the surface of turbulent fluids, 
Kraichnan's model fails in basic qualitative predictions
\cite{Boffetta04,Larkin}
\del{. This finding} has raised a major theoretical challenge
which remained open regardless of recent studies of the role of time correlations, 
cf.\ \added{\cite{Gustavsson2013,Dhanagare2014,Jurchishinova,Duplat2000,Falkovich2007}}.

\del{Long-term behavior of passive fields in smooth chaotic flows is defined by the Lagrangian Lyapunov exponents 
$\lambda_i$, $i=1\ldots d$, together with the exponents $\varkappa_i$ describing 
the variance of finite-time  Lyapunov exponents;
here $d$ denotes the topological dimension of the flow, and we assume ordering $\lambda_{k}\ge\lambda_{k+1}$. 
Indeed,} 
One can show (e.g.\ using the approaches of \cite{Balkovsky1999,Boldyrev,JK0}) that asymptotically, when the observation time becomes much greater than the flow correlation time,
the probability distributions of the 
\del{length, face height and volume of a material parallelepiped become lognormal, i.e.\ normal in logarithmic length scale. 
Both the median and variance of this Gaussian distribution grow linearly in time,
the median's speed \del{equals to the corresponding}
\added{is the}
Lyapunov exponent; half of the variance growth rate will be called as the \textit{Lyapunov exponent diffusivity} and denoted by $\varkappa_i$.
Only ``fat-tailed'' time correlations (e.g.\ due to 
near-wall stagnant regions, 
cf.\ \cite{Gouillart}) can invalidate this scenario.
Hence, if these exponents are equal for two flows, the stretching statistics is identical at the long-time limit.}
\added{
logarithms of the dimensions of a material parallelepiped become Gaussian.
Both the median and variance of such a Gaussian distribution grow linearly in time,
the median's speed 
is \textit{the Lyapunov exponent $\lambda_i$}; half of the variance growth rate will be called as the \textit{Lyapunov exponent diffusivity} and denoted by $\varkappa_i$. Here, $i=1$ refers to the length, $i=2$ --- to the face height,
and $i=3$ --- to the height of a material parallelepiped. Thus, the stretching statistics is fully described by the set of exponents $\lambda_i, \kappa_i$, $i=1\ldots d$ ($d$ denotes the flow dimensionality).
Only ``fat-tailed'' time correlations (e.g.\ due to 
near-wall stagnant regions \cite{Gouillart}) can invalidate this scenario.
}

One of the listed exponents can be 
equated to unity with a proper choice of time unit. Thus, what defines the dynamics of passive fields
are the ratios of the exponents.
So, the efficiency of small-scale kinematic magnetic dynamos \cite{Gruz} 
and the exponent of the statistical conservation law for Lagrangian pair dispersion \cite{Falkovich2013,Frishman}
depend on the ratio $\lambda_1/\varkappa_1$ ; 
multifractal spectra of scalar dissipation fields depend on 
$\lambda_d/\varkappa_d$ \cite{JK0}, the same applies to 
the dynamics of monomolecular reactions in compressible flows \cite{Kalda11};
the (multi)fractal properties of tracer fields in compressible 2-dimensional (2D) flows 
depend on $\lambda_2/\lambda_1$ \cite{Bec04,Boffetta04}.

The structure of this Letter is as follows.
For statistically isotropic homogeneous Kraichnan flows, the exponent ratios 
\del{can be expressed purely in terms of}
\added{are expressed via}
the flow compressibility and dimensionality \cite{Falk}; 
first we discuss different factors contributing to the failure of these expressions for real flows.
Further, \added{we introduce a class of model flows; based on this model in 2D geometry,} we obtain expressions for $\lambda_1/\varkappa_1$ and $\lambda_1/\lambda_2$ which exhibit non-universality depending not only on the 
Kubo number $K$ (ratio of the correlation time and \added{small-scale} eddy turnover time), but also on the statistics of the determinant of the 
velocity gradient tensor. \added{We use qualitative arguments to show that the results are robust and not limited to our model.} 
Based on the fact that for incompressible Kraichnan flows, $\lambda_1/\varkappa_1=d$, the ratio
$\lambda_1/\varkappa_1\equiv d_m$ will be called the \textit{mixing dimension of the flow}; we argue that for time-correlated flows, $d_m>d$, 
and discuss implications of this inequality.
Finally, we show that when exponents are averaged over a realistic ensemble of determinant values, the contradiction between 
the theory and experimental findings \cite{Boffetta04,Larkin} becomes resolved. 

\textit{Factor A: different correlation times for potential and solenoidal components.}
\del{In the case of partially compressible flows,} The mismatch between real and Kraichnan flows can be caused by the solenoidal 
and potential components (denoted as $\boldsymbol{v}_s$ and $\boldsymbol{v}_p$) of the Helmholtz 
decomposition of the velocity field $\boldsymbol{v}\equiv\boldsymbol{v}(\boldsymbol{r},t)$ having different correlation times 
$\tau_s$ and 
$\tau_p$, respectively.  
For instance in marine environment, the potential component of surfaces flow is associated with 
up- and down-welling, and is often persistent due to being caused by bathygraphy \cite{KaldaGiud}.
However, we can modify the Kraichnan model so as to cover the case  $\tau_s\ne\tau_p$
\del{First} if we 
\del{ascribe to the flows} \added{introduce} very small correlation times 
$\tau_s,\tau_p$. Then, for statistically isotropic homogeneous case with 
$\tau_s=\tau_p\equiv\tau $,
\begin{align*}
\lambda_i&=D_1\left\{d(d-2i+1)-2\wp [d+(d-2)i]\right\},\\
\varkappa_i&=D_1(d-1)(1+2\wp), \numberthis\label{Kr}
\end{align*}
where the compressibility $\wp\equiv{\cal F}_p/{\cal F}$
is defined via the mean squared Frobenius norms 
${\cal F}_p\equiv\langle\|\boldsymbol{\nabla v}_p\|^2\rangle=\mbox{$\langle(\boldsymbol \nabla \cdot\boldsymbol v)^2\rangle$}$,
and ${\cal F}\equiv\langle\|\boldsymbol{\nabla v}\|^2\rangle$; 
the factor $D_1=\frac{{\cal F}\tau}{d(d-1)(d+2)}$, where
$\tau=\frac 1{\cal F}\int_0^\infty\langle \partial_iv_j(\boldsymbol r,0) \partial_iv_j(\boldsymbol r,t)\rangle \mathrm{d} t$,
cf.\ \cite{Falk}.
If we substitute in this integral $\boldsymbol v=\boldsymbol v_p+\boldsymbol v_s$ and assume $\tau_s\ne\tau_p$,
Eq.~(\ref{Kr}) is generalized to
$\lambda_i={\cal F}_s\tau_s
\frac{d-2i+1}{(d-1)(d+2)}+{\cal F}_p\tau_p
\frac{d-4i}{d(d+2)}$, and $\varkappa_i={\cal F}_s\tau_s
\frac{1}{d(d+2)}+{\cal F}_p\tau_p
\frac{3}{d(d+2)}$,
\added{where ${\cal F}_s\equiv\langle\|\boldsymbol{\nabla v}_s\|^2\rangle$.}
Now it becomes clear that for the exponent ratios to depend only on $\wp$ and $d$, and for Eq.~(\ref{Kr}) to remain valid
with $D_1=\frac{{\cal F}_s\tau_s+{\cal F}_p\tau_p}{d(d-1)(d+2)}$, the compressibility needs to be 
defined as 
%184
\begin{equation}\label{compr}
\wp =\tilde\wp\equiv{\cal F}_p\tau_p/({\cal F}_s\tau_s+{\cal F}_p\tau_p).
\end{equation}

Our simulations with 2D sine flow \added{(velocity field is sinusoidal in space and a discretely updated random function in time,
cf.\ Supplementing Material \cite{Ainsaar_integral})} show that for Kubo number $K=\tau\sqrt{\cal F}\ll 1$, 
the generalized expression of compressibility $\tilde\wp$ describes adequately flows 
with $\tau_p\ne\tau_s$, see Fig.~1. However, for $K\sim 1$, a
considerable mismatch is observed, confirming the findings of ref.\ \cite{Boffetta04}.

\begin{figure}[tb]
\includegraphics{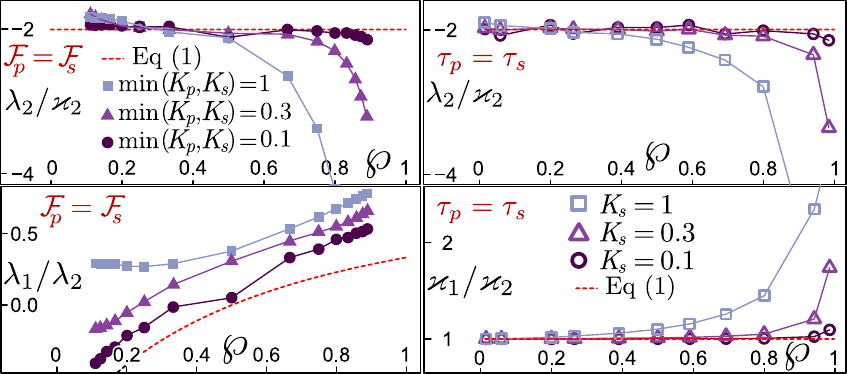}%
\caption{Lyapunov exponent ratios from simulations with sine flows 
are plotted versus the compressibility $\wp$ as defined by Eq.~(\ref{compr}). 
Left column: solenoidal and potential velocity components obey equal norms but different correlation times; %$\tau_p\ne\tau_s$;
right column: while $\tau_p=\tau_s$, norms are different. 
\label{mihklilt}}
\end{figure}

\textit{Factor B: correlations along Lagrangian trajectories due to particles embedded into material volumes.}
Hypersonic flows, if judged by velocity gradient, are characterized by a considerable compressibility; however,
the sum of Lyapunov exponents is strictly zero (limitlessly contracting volumes would imply limitlessly growing pressure fluctuations).
This violates Eq.~(\ref{Kr}) which predicts $\sum_i\lambda_i=-D_1d(d-1)(d+2)\wp$.
The mismatch is explained by long-term correlations. 
\del{If we want to apply the results based on Kraichnan  flows to such hypersonic flows, we need to match the stretching statistics and use $\wp=0$.}

\textit{Factor C: genuine finite correlation time effects.}
Even if the factor A is accounted for by using
generalized compressibility (\ref{compr}), and the factor B is negligible, time correlations can 
cause a considerable departure from Eq.~(\ref{Kr}).
We follow the approach developed in \cite{JK0,JK07PRL} and
apply multiplicatively the deformations achieved during a correlation time $\tau$ assuming that beyond a correlation time, 
the correlations are negligible. Then, infinitesimal material vectors $\boldsymbol{\D x}$ are transformed according to the tensor 
$\hat {\mathcal D}_t= \exp({\int_t^{t+\tau} \boldsymbol\nabla \boldsymbol{v}\D t'})$, $\boldsymbol{\D x}_{n\tau}=\prod_{j=0}^{n-1}\hat {\mathcal D}_{j\tau}\cdot\boldsymbol{\D x}_0$.

\textit{First, let us consider incompressible flows.}
In the case of 2D geometry, we need to study just the normalized \del{(to its initial length)} length of the vector 
$\ell_t\equiv |\boldsymbol{\D x}_{t}|/|\boldsymbol{\D x}_0|$. 
Indeed,
at the limit $t\to \infty$, $\lambda_1=\frac 1t\langle \ln\ell_t\rangle$ and $\varkappa_1=\frac 1{2t}\langle (\ln\ell_t-\lambda_1t)^2\rangle$.
\del{ 
If we consider a material parallelogram of normalized length $\ell_t$ and %normalized 
height $h_t$ then its surface area remains equal to unity, and
$\ln \ell_t\equiv-\ln h_t$; this leads us to }
\added{
A square built on $\boldsymbol{\D x}$ evolves into a parallelogram of normalized height $h_t$
and constant surface area, $\ell_th_t=1$, therefore}
$\lambda_2=\frac 1t\langle \ln h_t\rangle=-\lambda_1$ and $\varkappa_2=\varkappa_1$. 
Furthermore, the increments $\ln\ell_{t}-\ln\ell_{t-\tau}$ and $\ln\ell_{t+\tau}-\ln\ell_{t}$ are uncorrelated, hence
$\lambda_1\equiv\lim_{t\to\infty} \frac 1t\langle \ln\ell_t\rangle=\frac 1\tau\langle \ln\ell_\tau\rangle$ and 
similarly, $\varkappa_i=\frac 1{2\tau}\langle (\ln\ell_\tau-\lambda_1\tau)^2\rangle$. %So, it is enough to average over the correlation time $\tau$.
Averages $\langle \ln\ell_\tau\rangle$ and  $\langle (\ln\ell_\tau)^2\rangle$ will be calculated in two stages:
averaging over rotations of a fixed \del{transformation} tensor  $\hat {\mathcal D}_\tau\equiv\hat {\mathcal D}$ 
(using statistical isotropy), and averaging over an 
ensemble of matrices $\hat {\mathcal D}$.

For a fixed transformation tensor $\hat {\mathcal D}$, $\ell_\tau^2=\boldsymbol e^T\hat {\mathcal D}^T\hat {\mathcal D}\boldsymbol e$,
where $\boldsymbol e\equiv \D \boldsymbol x/|\D \boldsymbol x|$. 
\del{is a unit vector  parallel to $\D \boldsymbol x$.}
The (Green's) deformation tensor
$\hat {\mathcal D}^T \hat {\mathcal D}$ is symmetric and 
positive semidefinite, hence, it has non-negative eigenvalues $p^2$ and $q^2$ \added{($q=1/p$ if $\wp =0$),}  and is
diagonalizable via a rotation into $\diag(p^2, q^2)$. 
Thus,
$\ell_\tau^2 = p^2 \cos^2\alpha + q^2 \sin^2\alpha$, and 
\begin{align}
&\lambda_1\tau=\int_0^{\frac{\pi}{2}} \ln
% p^2 \cos^2 \alpha + q^2 \sin^2 \alpha 
\ell_\tau^2\,
\frac{\D\alpha}\pi
=
\ln \frac{p+q}{2} \text,
\label{int_ln}
\\
&\varkappa_1\tau=\int_0^{\frac{\pi}{2}} \left( 
%p^2 \cos^2 \alpha + q^2 \sin^2 \alpha 
\ln \ell_\tau^2
\right)^2  \frac{\D\alpha}{4\pi}-\frac{\lambda_1^2\tau^2}2
%\nonumber\\ &\quad {}
=
\frac 14 \operatorname{Li}_2\! \frac{(p-q)^2}{(p+q)^2}
\text,
\label{int_ln_square}
\end{align}
where $\operatorname{Li}_2 x = - \int_0^x \left[ \ln(1-u) \right] \, \frac{\D u}u$ is
the dilogarithm, and $p\ge q>0$. The integral of Eq.(\ref{int_ln}) has been tabulated in 
\cite[Eq.~2.6.38.4]{Prudnikov}, and derived in \cite{AinsaarKalda};
the one of Eq.~(\ref{int_ln_square}) is derived in the Supplementing Material \cite{Ainsaar_integral}.

\begin{figure}
\includegraphics{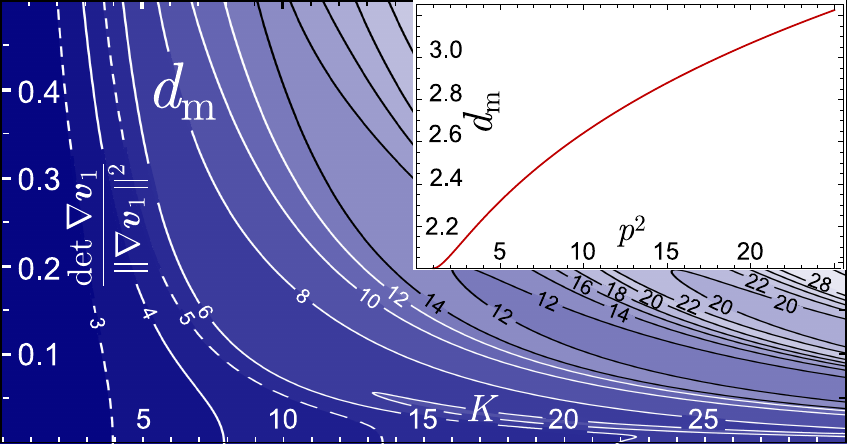}%
\caption{
Mixing dimension $d_m\equiv\lambda_1/\varkappa_1$ of 2D incompressible \added{model} flows \added{(with fixed values of $\frac{|\det\nabla\boldsymbol{v}_1|}{\|\nabla\boldsymbol{v}_1\|}$)} as a function of the 
determinant of the velocity gradient and the Kubo number $K\equiv\tau\|\boldsymbol{\nabla v}_1\|$;
\added{curves denote isolines and are labeled with the values of $d_m$.}
\del{dashed lines show local maxima of $d_m$ for fixed $K$ which are well approximated by
$\sin\left(K\sqrt{\det\nabla\boldsymbol{v}_1}/\|\nabla\boldsymbol{v}_1\|\right)=0$.}
Insert: $d_m$ versus the transformation tensor's singular value $p$.
\label{lambda-kappa}}
\end{figure}

Using Eqns.~(\ref{int_ln},\ref{int_ln_square}), we have plotted the mixing dimension as a function of the 
parameter $p$ (Fig.~\ref{lambda-kappa}, insert), which demonstrates non-universality. Indeed, 
we see that $d_m$ grows limitlessly with $p$, and $p$, in its term, can take arbitrarily large values 
if $K$ is large enough. \del{and $\det \boldsymbol{\nabla v}<0$.}
\added{In order to understand how $d_m$ depends on the velocity gradient statistics, we assume simplifyingly that $\boldsymbol{\nabla v}$ remains constant during one correlation time [$\boldsymbol{\nabla v}(t)\equiv \boldsymbol{\nabla v}_1$],
upon which it obtains a new random value. Real flows, of course, depend smoothly on time, but
$\boldsymbol{\nabla v}_1$ can be interpreted as such a time-averaged tensor 
$\boldsymbol{\nabla v}(t)$ that during one correlation time, it results in the same deformation tensor 
$\hat {\mathcal D}^T \hat {\mathcal D}$ as the real smooth-in-time flow.
}
Statistically homogeneous flows include both saddles ($\det \boldsymbol{\nabla v}<0$) and extrema ($\det \boldsymbol{\nabla v}>0$) of the streamfunction and $\left<\det \boldsymbol{\nabla v}\right>=0$. The smallest possible ensemble 
of tensors $\boldsymbol{\nabla v}$ satisfying this constraint includes two fixed tensors $\boldsymbol{\nabla v}_1$ and $\boldsymbol{\nabla v}_2$ with $\det \boldsymbol{\nabla v}_1+\det \boldsymbol{\nabla v}_2=0$;
the main graph of Fig.~\ref{lambda-kappa} shows the behavior of $d_m$ for such ensembles.
For $K\ll 1$, the values of $d_m$ remain close to (and larger than) the Kraichnan limit value $d=2$.
For $K\agt 1$, 
much larger values can be reached.

Let us discuss now the significance of the mixing dimension $d_m\equiv \lambda_1/\varkappa_1$.  
\del{For incompressible Kraichnan flows we need to have $d_m=d$.
Indeed, consider the evolution of the distribution $\sigma$ of inter-particle distance $r$ for two tracers, 
initially seeded at a small distance $r_0$, using logarithmic scale $\mu\equiv\ln r/r_0$. 
}
\added{First we show that for incompressible Kraichnan flows we need to have $d_m=d$.
Let $\mu\equiv\ln r/r_0$ denote the distance of tracer particles, initially seeded at a small distance $r_0$, using logarithmic scale.
}
Then, $\mu$ performs a biased random walk
with mean velocity $\lambda_1$ and diffusivity $\varkappa_1$; the step length is defined by the correlation time of the flow, cf.\ \cite{JK07PRL}.
Thus, the asymptotic-in-time evolution equation 
\added{for the distribution function $\sigma=\sigma(\mu,t)$}
is written as $\partial_t\sigma+\lambda_1\partial_\mu\sigma=\varkappa_1\partial^2_\mu\sigma$. 
The distance $r$ is limited by inter-molecular distances $r_{\min}$, hence we need to apply a reflective boundary condition 
at $\mu_{\min}=\ln r_{\min}/r_0$; 
\del{With such a boundary condition and after waiting long enough ($t\agt -\mu_{\min}/\lambda_1$), a quasi-stationary distribution $\sigma\propto \textrm e^{d_m\mu}$ is reached at small values of $\mu$; }
\added{this leads to a quasi-stationary distribution $\sigma\propto \textrm e^{d_m\mu}$ at small values of $\mu$
if we wait long enough ($t\agt |\mu_{\min}|/\lambda_1$).}
\del{this translates into cumulative distribution function in $r$-space being proportional to $r^{d_m}$.}
\added{Therefore, the cumulative distribution function in $r$-space is proportional to $r^{d_m}$.}

Now, consider the dye distribution around a  material point $P$, 
with a spot of dye being seeded initially at a small distance $r_0$ from it.
Equality $d=d_m$ implies a locally homogeneous mixing: the amount of dye inside 
a sphere around $P$ scales as the volume of the sphere.
Inequality $d_m<d$ is clearly impossible for incompressible flows as it would mean local clustering of the material particles.
Meanwhile, $d_m>d$ implies local inhomogeneity in mixing: 
with a high probability, a neighborhood of $P$ is left without dye. This scenario cannot be realized \added{(hence, $d=d_m$)} in the case of Kraichnan flows when
the material deformations 
are diffusive in nature.
Indeed, at small time-scales, diffusion dominates over the mean drift in the dynamics of $\mu\equiv\ln r/r_0$, and therefore, 
the distance $\Delta \mu=\ln r_0/r_{\min}$ to the reflection point of $\mu$ is traveled diffusively. 
This ensures that all those dye particles which start from the vicinity of $P$ will also visit the smallest 
neighborhood of $P$, i.e.\ there is a locally homogeneous mixing described by $\sigma\propto r^d$.
\del{, hence $d_m=d$.}
Meanwhile, for real flows, diffusive behavior is reached only asymptotically,  at time scales larger than $\tau$. 
So, there is a chance that with few initial steps, the dye particle ``runs away'' from the point $P$.
At long time scales, drift dominates over 
diffusion; thus, if the dye particle managed to evade the point $P$ at the early stage, 
it will avoid it forever. 
\del{(This discussion can be  
generalized to the case of non-smooth streamlines within inertial range of turbulence.)}

\textit{The case of compressible fields} has been studied preliminarily in a technical report \cite{AinsaarKalda}; 
here we use our model to explain the 
challenging findings of Ref.~\cite{Boffetta04}
\added{and provide a qualitative explanation of this phenomenon.}
To keep the overall area of the basin constant, 
both divergent and convergent regions need to be present simultaneously. 
We'll use the same ensemble for $\boldsymbol{\nabla v}$ as before, 
with an additional requirement $\boldsymbol\nabla\cdot\boldsymbol v_1=-\boldsymbol\nabla\cdot\boldsymbol v_2\equiv\delta>0$.
We assume simplifyingly that the convergent (`C') and divergent (`D') domains are separated with a sharp boundary which remains
constant throughout the full correlation time $\tau$, upon which new random domain breakup and tensor rotation angles are coined.
\del{While in reality, divergence is a smooth function of coordinates and time,
{\em (a)} our artificial field can still prove non-universality of mixing;
{\em (b)} it appears to provide results consistent with realistic simulations, hence we argue that it captures the essential features of real flows.}
\added{While this may seem to be an unrealistic simplification, we'll argue later that 
our model captures essential features of realistic flows.
}

During one correlation time, some of the tracer particles are carried from domain `D' to domain `C', let us consider such particles which cross the domain boundary at the moment $t$ (measured from when the current domains were established), henceforth referred to as the $t$-particles. 
Eq.~(\ref{int_ln}) allows us to average the Lyapunov exponent over $t$-particles, 
$\left<\lambda_1\right>_{t\textrm{-p}}\equiv \Lambda(t)=
\frac 1\tau\ln\left(\frac{p_1+q_1}2\frac{p_2+q_2}2\right)$; here
$p_1$ and $q_1$ are the singular values of 
$\textrm e^{\boldsymbol\nabla \boldsymbol{v}_1(\tau-t)}$, and 
$p_2$ and $q_2$ are that of $\textrm e^{\boldsymbol\nabla \boldsymbol{v}_2t}$.

Within region `D', surface areas grow exponentially $A\propto \textrm{e}^{\delta t}$, hence, the tracer density 
falls as $\textrm{e}^{-\delta t}$. 
Thus, the probability for a random particle to exit the region between $t$ and $t+\D t$ is given by $\D p=\textrm{e}^{-\delta t}\delta\D t$. Now we can find the overall Lyapunov exponent by averaging $\Lambda(t)$ over all the tracer particles, 
$\lambda_1=\left<\Lambda(t)\right>=\frac 12\left[\int_0^\tau\Lambda(t)\delta \textrm{e}^{-\delta t}\D t+\Lambda(\tau)\textrm{e}^{-\delta t}+\Lambda(0)\right]$. 
Here, the second and third terms correspond to those particles which remain for the whole period $\tau$ within `D'  and `C', respectively.
The area growth exponent 
$\lambda_A=\lambda_1+\lambda_2$ is calculated similarly: 
in region `D', $\lambda_A=\delta$, and in region `C', $\lambda_A=-\delta$. 
At the position of a $t$-particle, the area growth factor is $\textrm{e}^{\delta t -\delta (\tau-t)}$; 
once we average over all the values of $t\in\left[0,\tau\right]$, we end up with
$\lambda_A=\frac 1\tau\left(1-\delta \tau-e^{-\delta\tau}\right)$. Finally,
\del{the smallest Lyapunov exponent is expressed as}
\added{we can express}
$\lambda_2=\lambda_A-\lambda_1$.

\begin{figure}
\includegraphics{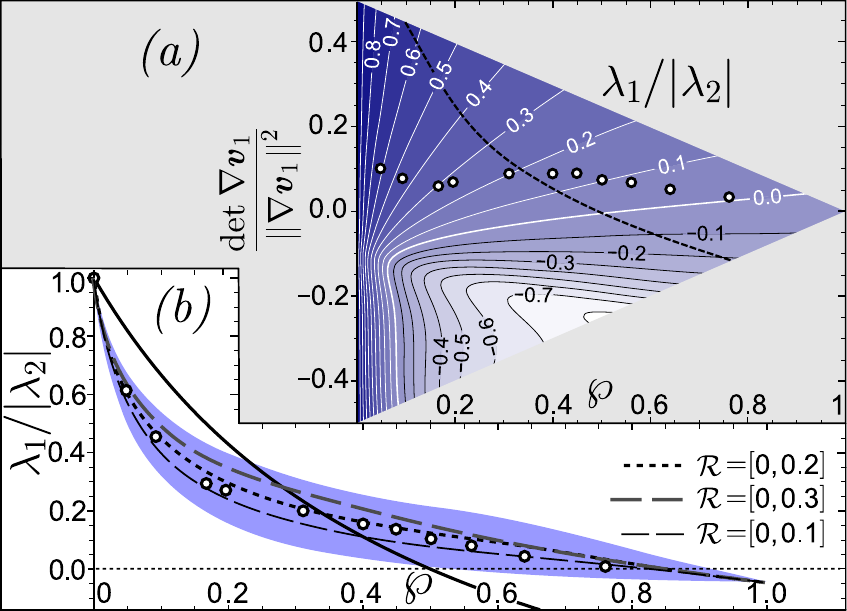}
\caption{\textit{(a)} $\lambda_1/|\lambda_2|$ as a function of $\wp$ and $\frac{\det \nabla \boldsymbol{v}_1}{\|\nabla \boldsymbol{v}_1\|^2}$ 
for $K=9$; \added{numbers indicate the isoline values} 
\del{the values range from -0.73 (darkest) to 1 (brightest)}
Graph area is triangular because 
for any $\nabla \boldsymbol{v}_1$, $\frac{2|\det \nabla \boldsymbol{v}_1|}{\|\nabla \boldsymbol{v}_1\|^2}\le 1-\wp$.
Dots show the intersection points of this graph with the data of Ref.~\cite{Boffetta04}, \added{and dashed curve --- with the theoretical dependence for Kraichnan flows}.
\textit{(b)} 
Shaded region shows the range of values of $\lambda_1/|\lambda_2|$ 
for $\frac{\det \nabla \boldsymbol{v}_1}{\|\nabla \boldsymbol{v}_1\|^2}\in [0,0.2]$;
\added{solid curve: theoretical dependence for Kraichnan flows; 
dots: data of Ref.~\cite{Boffetta04};
dashed lines:  the exponents are averaged over $\frac{\det \nabla \boldsymbol{v}_1}{\|\nabla \boldsymbol{v}_1\|^2}\in {\cal R}$, the ranges
${\cal R}$ being shown by the legend.}
\label{fig:contour}
\label{fig:clouds}}
\end{figure}

We used our expressions 
to compute $\frac{\lambda_1}{|\lambda_2|}$ as a function of the model parameters; in Fig.~\ref{fig:clouds}(a), the 
results are depicted for $K=9$. 
\added{Dashed curve marks the intersection of this graph with the Kraichnan-limit-dependence 
$\frac{\lambda_1}{|\lambda_2|}=\frac{1-2\wp}{1+2\wp}$ and
divides the plane into two regions: at the left-bottom part, $\frac{\lambda_1}{|\lambda_2|}$ is decreased due to time-correlations, and at the right-top part, the effect is opposite.

At small compressibilities [$\wp<0.1$ for Fig.~\ref{fig:clouds}(a)], 
$\frac{\lambda_1}{|\lambda_2|}$ is always reduced;
we show that this is a robust feature, not bound to our model. 
Consider $\lambda_1$
as a function of $K$ for flows with a fixed probability 
distribution of  $\boldsymbol{\nabla v}$:  
Eq.~(\ref{Kr}) holds while $K\ll 1$, and gives us estimate 
$\lambda_1\sim\tau {\cal F}=K\sqrt{\cal F}$.
By $K\gg 1$, a saturation value is reached, which is estimated as the stretching rate near saddles,  $\lambda_1\sim\sqrt {\cal F}$. 
Exponent $\lambda_A$ behaves similarly, but saturates later:
for material areas, the Kraichan approximation holds as long as
the area change during $\tau$ remains small, $\tau\boldsymbol\nabla \cdot \boldsymbol v= K\sqrt\wp\ll 1$, and yields 
$\lambda_A\sim -K\wp\sqrt{\cal F}$; for  $K\sqrt\wp\gg 1$,
saturation $\lambda_A\sim -|\boldsymbol\nabla \cdot \boldsymbol v|$ is reached.
If $K\agt 1$ and $\wp\alt K^{-2}$, 
saturation is reached for $\lambda_1$, but not for $\lambda_A$, i.e.\  
$\frac{|\lambda_2|}{\lambda_1}=1-\frac{\lambda_A}{\lambda_1}\sim 1+K\wp$.
This means that larger values of $K$ lead to faster growth of $\frac{|\lambda_2|}{\lambda_1}$ with $\wp$.

%\wp-4*det
At larger compressibilities [$\wp\agt 0.3$ for Fig.~\ref{fig:clouds}(a)], 
it can be approximately said that $\lambda_1>0$ for $\det\boldsymbol\nabla\boldsymbol{v}_1>0$ and $\lambda_1<0$ for $\det\nabla\boldsymbol{v}_1<0$. This is also a robust feature.
Indeed, at the limit $K\gg 1$, the tracers stay mostly near the most convergent areas, hence
\textit{$\lambda_1$ is dominated by the stretching statistics in the regions with smallest values of $\boldsymbol\nabla\cdot\boldsymbol{v}$}. Therefore, we can assume that along Lagrangian trajectories,
$\boldsymbol\nabla\boldsymbol{v}$ remains almost constant for a long period of time; 
under the assumption of constant $\boldsymbol\nabla\boldsymbol{v}$, one can show that $2\lambda_1=\boldsymbol\nabla\cdot\boldsymbol{v}+\operatorname{Re}\sqrt{(\boldsymbol\nabla\cdot\boldsymbol{v})^2-4\det\boldsymbol\nabla\boldsymbol{v}}$; hence, with $\boldsymbol\nabla\cdot\boldsymbol{v}<0$, the sign of
$\lambda_1$ is opposite to that of $\det\boldsymbol\nabla\boldsymbol{v}$.
In the case of our model flow, this corresponds to $\operatorname{sgn} \lambda_1=\operatorname{sgn} \det\boldsymbol\nabla\boldsymbol{v}_1$.
}

Boffetta et al.\cite{Boffetta04} found the Lyapunov dimension $d_L=1+\lambda_1/|\lambda_2|$
as a function of \del{the compressibility} $\wp$ for  
realistic compressible flows with numerically increased  
correlation time at $K\approx 9$.
%(assuming $\lambda_1>0)$ 
To compare our theory with Ref.~\cite{Boffetta04}, Lyapunov exponents need to be averaged
over a realistic ensemble of velocity
gradients. We were unable to find data for free-slip liquid interfaces; thus, the analysis is
based on the data for  2D
slices of an incompressible 3D velocity field \cite{Cardesa13,Rabey15}; these papers report 
the joint probability density of the trace (there
called ``$-p$'') and determinant (``$q$'') of the velocity gradient.
There is a small difference in the normalization of $\det\nabla\boldsymbol{v}$:
Refs.~\cite{Cardesa13,Rabey15} normalize by  
$\langle (\nabla \times \boldsymbol{v})_z^2 \rangle$.
\del{
From the fact that $\langle\det\nabla\boldsymbol{v}\rangle=0$ together with the fact that before
taking the slice their 3D velocity field $\boldsymbol{v}_{\mathrm{3D}}$ was incompressible ($\nabla \cdot
\boldsymbol{v}_{\mathrm{3D}} = 0$), it can be seen that our normalization factor
has only an extra $\langle (\partial_z v_z)^2 \rangle$.
Assuming isotropy, this means that our divisor }
\added{
Based on the fact that before
taking the slice their 3D velocity field $\boldsymbol{v}_{\mathrm{3D}}$ was incompressible, and assuming isotropy,
it can be shown that our normalization factor}
is $\frac 43$ times %($15\%$)
larger. Keeping this in mind, Fig.~3 of Ref \cite{Cardesa13} tells us that 
within regions `D' with $p < 0$, the values of 
$\frac{\det \nabla \boldsymbol{v}_1}{\|\nabla \boldsymbol{v}_1\|^2}$ 
remain mostly between 0 and 0.2. When we average the Lyapunov exponents over this  range, 
our model provides a good match to the data of Ref.~\cite{Boffetta04}, see Fig.~\ref{fig:contour} (bottom).

In conclusion, we have expressed the ratios of Lyapunov exponents and their diffusivities in terms of velocity 
gradient statistics, and introduced the mixing dimension $d_m$ as a characteristic of a flow. 
Those existing theories which are based on the values of these ratios (for instance\ \cite{Gruz,Bec04})
become directly applicable to real time-correlated flows.
\del{In the case of compressible flows, the Kraichnan flow compressibility can be used as a free parameter 
to obtain a better modeling of real flows. }
We also pinpoint a major qualitative difference: \del{between real  and Kraichnan flows:}
while open Kraichnan flows are characterized by locally homogeneous mixing (once we wait long enough, dye density will be 
distributed evenly over a neighborhood of a fixed material point), in the case of time-correlated flows, the mixing is locally inhomogeneous.
This difference stems from the fact that for Kraichnan flows, $d_m$ equals to the integer-valued topological dimension $d$, 
and cannot perfectly match real flows with fractional $d_m$.
\del{The degree of mixing inhomogeneity is characterized by the difference $d_m-d$.}
Such an  inhomogeneity has major implications, in particular for the rate of bimolecular reactions with initially separated reagents (cf.\ \cite{Vassilicos2007,Ait-Chaalal}), 
for the mixing of different dyes (cf.\ recent experimental study \cite{Kree13}), and for the efficiency of kinematic magnetic dynamos \cite{Gruz}.
The local inhomogeneity is also likely the reason behind non-universality of mixing with respect to the details of tracer injection \cite{Gotoh2015}. 
\del{Finally, we have explained the findings of Ref.~\cite{Boffetta04}; the qualitative explanation 
stems from two observations: at the limit of large Kubo numbers, (a) the Lyapunov exponents are determined by 
the strain rate near the most convergent areas of compressible flows; (b) for small compressibilities,
the sum of Lyapunov exponents grows with the correlation time even if individual Lyapunov exponents have reached saturation.}

The support of  the EU Regional Development Fund Centre of Excellence TK124 (CENS) is acknowledged.

\newpage
\bibliographystyle{apsrev4-1}
\bibliography{diffus}

\end{document}